\documentclass[twocolumn,showpacs,prl]{revtex4}%
\usepackage{graphicx}
\usepackage{dcolumn}
\usepackage{bm}
\usepackage{amsmath}
\usepackage{amsfonts}
\usepackage{amssymb}%
\begin{document}
\preprint{ }
\title{$\nu=5/2$ Fractional Quantum Hall State in Low-Mobility Electron Systems:\\Different Roles of Disorder}
\author{Gerardo Gamez}
\author{Koji Muraki}
\affiliation{{\footnotesize NTT Basic Research Laboratories, NTT Corporation, 3-1
Morinosato-Wakamiya, Atsugi 243-0198, Japan}}
\date{\today }

\begin{abstract}
We report the observation of a fully developed fractional quantum Hall state at $\nu=5/2$ 
in GaAs/Al$_{x}$Ga$_{1-x}$As quantum wells with mobility well below $10^{7}$\thinspace cm$^{2}$/Vs. 
This is achieved either by strong illumination or reducing the barrier Al composition without illumination. 
We explain both results in terms of screening of the ionized remote impurity (RI) potential by nearby neutral shallow donors. 
Despite the dramatic improvement in the transport features, the energy gap $\Delta_{5/2}$ is limited to a rather small
value ($\sim100$ mK), 
which indicates that once the RI potential is well screened and the $5/2$ state
emerges, the size of $\Delta_{5/2}$ is limited by the mobility, i.e., by background impurities.
\end{abstract}

%\date{June 29, 2004} % It is always \today, today, but any date may be explicitly specified

%It is always \today, today, but any date may be explicitly specified


\maketitle

~The fractional quantum Hall (FQH) state at even-denominator filling factor $\nu=5/2$ 
in the $N=1$ first-excited Landau level (LL), 
whose origin has remained enigmatic 
since its discovery over 20 years ago~\cite{Willett}, 
is currently the focus of extensive studies. 
This is mainly because it constitutes a prime candidate for a non-Abelian state of
matter~\cite{Stern} and is considered as a potential platform for
implementing topological quantum computation~\cite{Nayak}. Experimental
access to the $\nu=5/2$ state, however, has been severely limited by the
extraordinary requirements imposed on the sample quality by the small energy scale
involved. Indeed, it was only after a sample with very high mobility of
$\mu=1.7\times10^{7}$~cm$^{2}$/Vs became available that a fully developed
$\nu=5/2$ state with exact quantization was demonstrated~\cite{Pan 1999}.

Disorder in remote-doped two-dimensional electron systems (2DESs) has two main 
sources: ionized remote impurities (RIs) and background impurities
(BIs)~\cite{Hwang}. After nontrivial technological advances in molecular-beam
epitaxy reducing BIs to achieve $\mu$ $\sim1.7\times10^{7}$\thinspace
cm$^{2}$/Vs, the ultra-high-mobility regime of $\mu\sim3\times10^{7}%
$\thinspace cm$^{2}$/Vs has become accessible by replacing the conventional
one-side-doped single heterostructure by a quantum well (QW) doped from both
sides~\cite{Pfeiffer 2003}. This allows one to place RIs further away from the
2DES while keeping the same electron density $n_{s}$. For such structures
with a typical setback distance of $100$~nm, the contribution of RI scattering
to $\mu$ is minor and $\mu$ is dominated by BI scattering \cite{Hwang}. The
reduced disorder has lead to the emergence of new correlated states in~the
$N=1$ LL~\cite{Eisenstein 2002,Xia,Kumar}. At the same time, the $\nu=5/2$ 
energy gap $\Delta_{5/2}$ has been observed to increase~\cite{Pan2,Choi} 
and effects of disorder on $\Delta_{5/2}$
have been discussed~\cite{Pan2,Choi,Dean,Nuebler}. Although it is widely believed that
$\mu$ $\geq10^{7}$\thinspace cm$^{2}$/Vs is necessary for observing a
well-developed $5/2$ state, the exact criteria dictating its emergence and
the mechanism limiting $\Delta_{5/2}$ are still unknown.

In this Letter, we demonstrate that a fully developed $\nu=5/2$ state can be
observed in GaAs/Al$_{x}$Ga$_{1-x}$As QWs with $\mu<10^{7}$\thinspace cm$^{2}%
$/Vs, without the need for illumination or special doping schemes
\cite{Friedland,Umansky}. We first show that, by strong illumination, a fully
developed $\nu=5/2$ state can be established in a sample with $x=0.34$ and 
$\mu$ as low as $4.8\times10^{6}$\thinspace cm$^{2}$/Vs. Furthermore, we show that, 
instead of illumination, reducing the barrier Al composition systematically improves
the FQH features, leading to a
dramatic emergence of a fully developed $5/2$ state at $x=0.25$. We explain
both results in terms of screening of RI potential by nearby neutral shallow
donors. Activation measurements reveal different roles of disorder in dictating the
emergence of the $5/2$ state and limiting $\Delta_{5/2}$.

We studied samples with a standard structure, $30$-nm-wide
GaAs/Al$_{x}$Ga$_{1-x}$As QWs modulation doped with Si from both sides at
setback distances of 100~nm above and 120~nm below the QW. We employed
conventional delta doping in the Al$_{x}$Ga$_{1-x}$As alloy. A series of 
samples with different $x$ and Si sheet doping concentrations ($N_{\mathrm{Si}}$) 
were grown. Here, $x$ was determined from low-temperature photoluminescence using the
energy of the exciton transition from the Al$_{x}$Ga$_{1-x}$As
barrier~\cite{Bosio}. Transport experiments were performed on $4$-mm square
specimens with InSn (50:50) contacts diffused at each corner. A standard
lock-in technique with excitation current of 20~nA and frequency ranging from
3 to 17~Hz was used. The samples were cooled in the mixing chamber of a
dilution refrigerator with a base temperature well below 20~mK.

We first show effects of illumination on the FQH features in the range
$2<\nu<3$. Figure~1(a) shows the longitudinal ($R_{xx}$) and Hall ($R_{xy}$)
resistances of a sample with $x=0.34$ and $N_{\mathrm{Si}}=2\times10^{12}%
$\thinspace cm$^{-2}$, taken without illumination. As expected for the
relatively low $\mu$ of this sample ($\mu=4\times10^{6}$\thinspace cm$^{2}$/Vs
at $n_{s}=2.48\times10^{11}$\thinspace cm$^{-2}$), only a poorly developed
minimum at $\nu=5/2$ is observed. $R_{xx}$ at $\nu=5/2$ is not thermally activated and the minimum
appears only as a result of the flanks that rise at low
temperatures~\cite{Willett,Gammel}. As the sample is successively illuminated
with a red LED and $n_{s}$ is increased, the visibility of the FQH features continues to
improve until $n_{s}$ reaches $3.60\times10^{11}$\thinspace cm$^{-2}$ with
$\mu=4.8\times10^{6}$\thinspace cm$^{2}$/Vs, where it saturates and no longer
increases upon further illumination at moderate intensity. Under these
conditions, clear $R_{xx}$ minima become visible at $\nu=8/3$ and $7/3$ as
well as at $\nu=5/2$ [Fig.\thinspace1(b)]. This level of improvement achieved
with moderate illumination (i.e., LED current of a few mA) is common and is
believed to be a consequence of the larger Coulomb energy and improved $\mu$,
both reflecting the higher $n_{s}$. However, the observed FQH features are 
not yet fully developed;
the $R_{xx}$ minima do not tend to zero and the plateaux in $R_{xy}$ are not
entirely defined even at the lowest temperature ($T\sim10$~mK).

\begin{figure}[b]
\centering
\includegraphics[width=75mm]{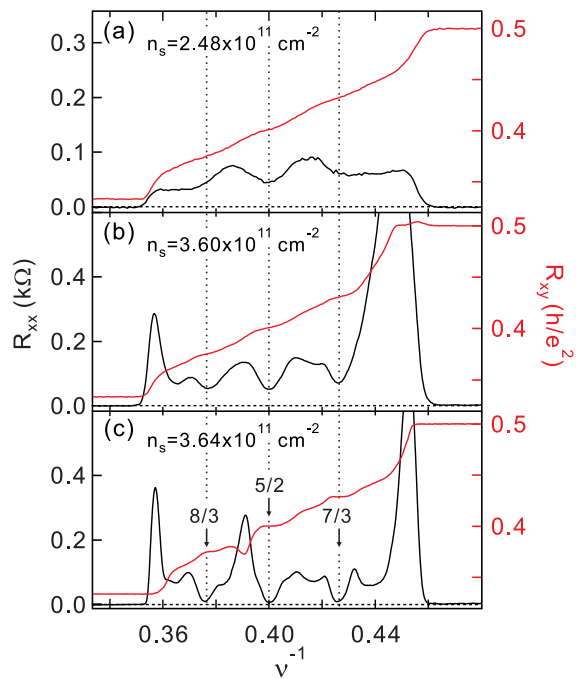}
\caption{(color online). $R_{xx}$ and $R_{xy}$ of the $x=0.34$ sample.
(a) Before illumination, (b) after moderate illumination, and (c) after strong 
illumination. $T$ is well below $20$ mK. All data were taken with the same 
contact configurations.}
\label{fig1}
\end{figure}

A dramatic improvement is observed when the sample is further illuminated at
much higher intensity (i.e., LED current of $\sim15$~mA) [Fig.\thinspace1(c)].
Now the FQH states at $\nu=5/2$, $8/3$, and $7/3$ are fully developed, with
$R_{xx}$ going to zero and $R_{xy}$ showing well-developed plateaux.
Temperature-dependent measurements demonstrate that $R_{xx}$ at $\nu=5/2$
follows an activated behavior $R_{xx}\propto\exp(-\Delta/2T)$, from which an
energy gap of $\Delta_{5/2}=93~$mK can be estimated. Note that the above
scenario, commonly used to account for the improvement in the FQH features by
illumination, does not apply to the present case, because $n_{s}$ increased 
only by $1\%$, and accordingly, $\mu$ barely changed with the strong illumination. 
These observations clearly demonstrate that the type of disorder that limits the mobility and the
one that governs the emergence of a fully developed $5/2$ state are distinct.
Although a rather poor correlation between $\mu$ and the appearance of the $5/2$
state was previously noted in the ultra-high-$\mu$ regime ($\gtrsim 3\times10^{7}$\thinspace cm$^{2}$/Vs)~\cite{Umansky}, 
the value of $\mu$ here is far outside that range and, indeed, lower than any of those at which a
fully developed $\nu=5/2$ state has ever been reported.

To clarify the mechanism underlying the dramatic improvement in the FQH
features, we first note the fact that Si donors in Al$_{x}$Ga$_{1-x}$As
($0.22<x<0.4$) can exist in two different states: a substitutional hydrogenic
shallow donor state and a deep center called a \textquotedblleft DX
center\textquotedblright\ accompanying lattice relaxation \cite{Mooney}.
When cooled in the dark, almost all Si donors become DX centers for
$0.3\lesssim x<0.4$, which can be transformed into shallow donors by
illumination at low $T$. This results in an increase in $n_{s}$ because
shallow donors have a larger energy offset with respect to the QW state in
GaAs. A second important fact is that $N_{\mathrm{Si}}\gg n_{s}$, so that
the saturation of $n_{s}$ does not necessarily imply that all DX centers have
been transformed into shallow donors. When placed in a slowly varying
potential created by ionized donors, shallow donors ($d^{0}$) may become
polarized by displacing their electronic wave functions, thereby screening
the disorder potential. The essence of our scenario is that while DX
centers have almost no screening capability due to their strongly localized
wave functions, shallow donors may provide good screening due to their much
more extended wave functions. Hence, the dramatic improvement in the FQH
features after strong illumination could be explained by the enhanced
screening provided by an increased number of $d^{0}$.

\begin{figure}[b]
\centering
\includegraphics[width=85mm]{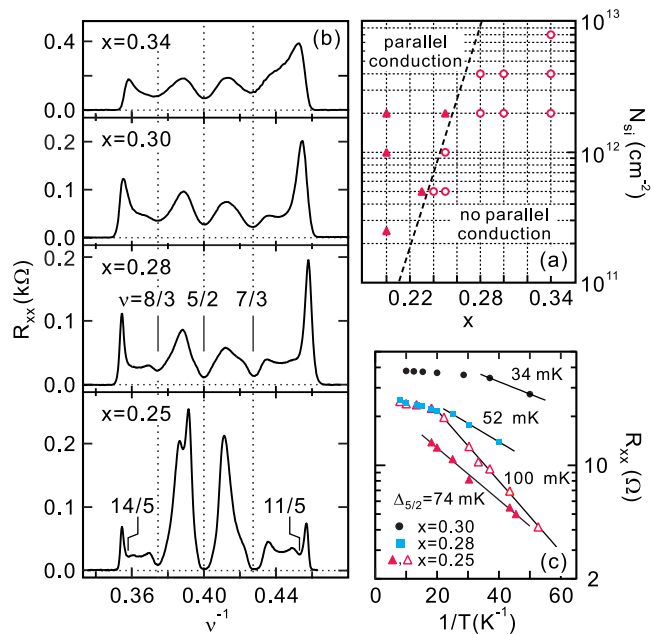}
\caption{(color online). (a) Mapping of samples in the $x$-$N_{Si}$
plane indicating the presence or absence of parallel conduction
without illumination. (b) $R_{xx}$ taken without illumination
from samples with different $x$ ($N_{\mathrm{Si}}$ = $1$-$2\times10^{12}$ cm$^{-2}$). 
From top to bottom, $n_{s}=2.48, 2.57, 2.73$, and $2.65\times10^{11}$ cm$^{-2}$. 
$T$ is well below $20$ mK. For $x$=$0.34$, data taken with a different
configuration than in Fig. 1(a) are shown. (c) Arrhenius plot
of $R_{xx}$ at $\nu=5/2$ for samples with different $x$. Open triangles 
show results for the sample shown in Fig.~3(b).
Lines indicate
slopes corresponding to $\Delta_{5/2}=100, 74, 52$, and $34$ mK.}
\label{fig2}
\end{figure}

DX centers lie deep in the band gap of Al$_{x}$Ga$_{1-x}$As for $x\sim0.4$ and
gradually approach the conduction band bottom with decreasing $x$, being taken
over by the hydrogenic donor level for $x<$ $0.22$~\cite{Mooney}. Hence, by
changing $x$ in the barrier, one expects that the donor state can be varied in
a more controlled manner than is possible by illumination. Figure 2(a) maps the samples we
investigated for this purpose in the $x$-$N_{\mathrm{Si}}$ plane. As $x$ is
decreased for a given $N_{\mathrm{Si}}$, even without illumination, parallel
conduction sets in, whose onset depends critically on $x$ and $N_{\mathrm{Si}%
}$. The onset of parallel conduction provides a measure of the wave-function
overlap between neighboring donors, which reflects the average donor distance
($\propto N_{\mathrm{Si}}^{-1/2}$) and the spatial extent of individual donor
wave functions. The results shown in Fig.~2(a) reveal that the donor state
drastically changes with $x$, even in the range of $x>0.22$ where DX centers
are believed to be mostly relevant \cite{Negative-U}.

Figure\thinspace2(b) depicts $R_{xx}$ taken without illumination from four
samples with different $x$ ranging from $0.34$ to $0.25$. These samples have
the same $N_{\mathrm{Si}}=2\times10^{12}$\thinspace cm$^{-2}$ except the one
with $x=0.25$, for which $N_{\mathrm{Si}}$ was reduced to $1\times10^{12}%
$\thinspace cm$^{-2}$ ($\equiv N_{0}$) to avoid parallel conduction
[Fig.\thinspace2(a)]. The data clearly demonstrate a systematic improvement of
the FQH features with decreasing $x$. The most dramatic change occurs when $x$
is decreased from $0.28$ to $0.25$, where the FQH states at $\nu=5/2$, $8/3$,
and $7/3$ become fully developed. The very low disorder in the $x$ $=$ $0.25$
sample is also seen from the emerging FQH features at $\nu=11/5$ and $14/5$
and from an additional minimum between $\nu=5/2$ and $8/3$, which is a precursor 
to the re-entrant insulating state \cite{Eisenstein 2002,Xia}. Note that all four
samples have similar $n_{s}$ of $(2.61\pm0.13)\times10^{11}$\thinspace
cm$^{-2}$, while $\mu$ varies as $4.0\,$, $5.0$, $5.5$, and $6.4\times10^{6}%
$\thinspace cm$^{2}$/Vs for $x$ $=$ $0.34$, $0.30$, $0.28$, and $0.25$. This
change in $\mu$ is considered to reflect the BI concentration in the Al$_{x}%
$Ga$_{1-x}$As barriers, which generally tends to decrease with decreasing $x$.
Figure\thinspace2(c) shows activated behavior at $\nu=5/2$, from which
$\Delta_{5/2}=74$ mK is estimated for the $x$ $=$ $0.25$ sample. Samples with
$x$ $=$ $0.28$ and $0.30$ also show activated behavior, but with smaller $\Delta
_{5/2}$.

As we already discussed, $\mu$ cannot be the reason for the dramatic change in
the FQH features with $x$. Rather, we note that the $x=0.25$ sample is close
to the onset of parallel conduction, which lies between $N_{\mathrm{Si}}%
=N_{0}$ and $2N_{0}$ for $x=0.25$ [Fig.~2(a)]. The spatial extent of the donor
wave functions at $x=0.25$ can be estimated from the average
donor distance, $N_{\mathrm{Si}}{}^{-1/2}=7$-$10$~nm. For $x=0.28\,$, on the
other hand, parallel conduction does not occur even when $N_{\mathrm{Si}}$ is
increased to $4N_{0}$ [Fig.~2(a)], indicating that the donor wave function is
much more localized~\cite{Number}. These results lend strong
support for our scenario that the screening of RI potential by nearby donor
electrons is playing an essential role in the emergence of a fully developed
$5/2$ state~\cite{Cluster}.

\begin{figure}[b]
\centering
\includegraphics[width=80mm]{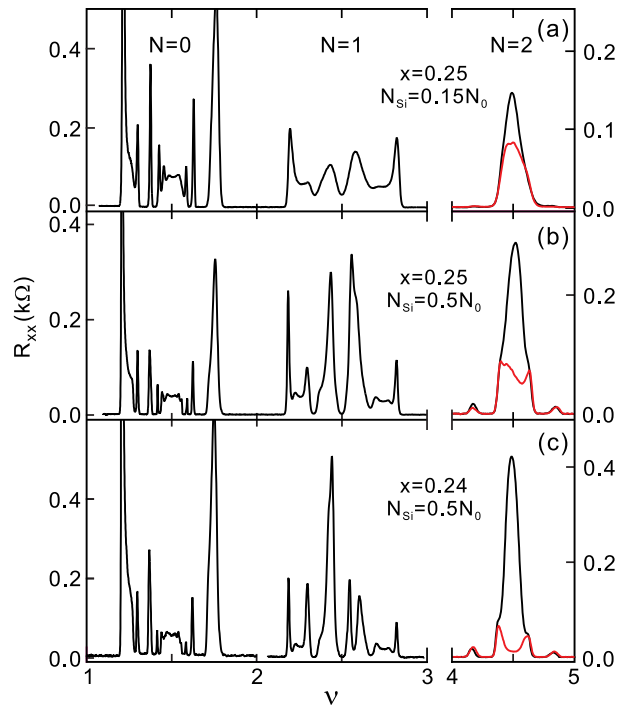}
\caption{(color online). $R_{xx}$ taken without illumination from
samples with different sets of $x$ and $N_{\mathrm{Si}}$. (a) $0.25$ and $0.15N_{0}$, 
(b) $0.25$ and $0.5N_{0}$, and (c) $0.24$ and $0.5N_{0}$ ($N_{0}=10^{12}$ cm$^{-2}$). 
From top to bottom, $n_{s}=2.43, 2.64, 2.66\times10^{11}$\,cm$^{-2}$ and $\mu=6.4, 7.6, 7.2\times10^{6}$\,cm$^{2}$/Vs. 
$T$ is well bellow $20$ mK for $\nu\leq3$, while $40$ mK for $\nu\geq4$.
The two curves for $\nu\geq4$ show $R_{xx}$ measured in the two orthogonal current directions.}
\label{fig3}
\end{figure}

Further insights are provided by examining the
dependence on the doping density. Figures\thinspace3(a) and 3(b) compare two
samples with the same $x=0.25$ but with lower $N_{\mathrm{Si}}$ of (a)
$0.15N_{0}$ and (b) $0.5N_{0}$. These samples have similar $n_{s}$ despite the
largely different $N_{\mathrm{Si}}$~\cite{Extra doping}. We show $R_{xx}$ in
three representative $\nu$ ranges to compare the influence of $N_{\mathrm{Si}}$
on the electronic
states in different LLs. For $N_{\mathrm{Si}}=0.5N_{0}$,
the $\nu=5/2$ state is equally well developed as in the case for
$N_{\mathrm{Si}}=N_{0}$. This is also confirmed by an activation measurement,
which shows an even larger gap of $\Delta_{5/2}=$ $100$ mK [Fig.~2(c)]. Further reducing
$N_{\mathrm{Si}}$ to $0.15N_{0}$, however, significantly deteriorates the
visibility of the $5/2$ state. We note that $N_{\mathrm{Si}}=$
$0.15N_{0}$ is close to $n_{s}/2$, which implies that
nearly all donors are ionized and so the data represent the case without
screening by donor electrons or correlation in the donor charge
distribution~\cite{Buks,Correlation}. Our result for
$N_{\mathrm{Si}}=$ $0.5N_{0}$ shows that, when the donor wave function is
sufficiently extended, a donor-electron density of
$N_{\mathrm{Si}}-n_{s}/2=0.37N_{0}$, which is only three times the
ionized-donor density $\sim n_{s}/2=0.13N_{0}$, is sufficient to 
establish a fully developed $5/2$ state.

\begin{figure}[t]
\centering
\includegraphics[width=85mm]{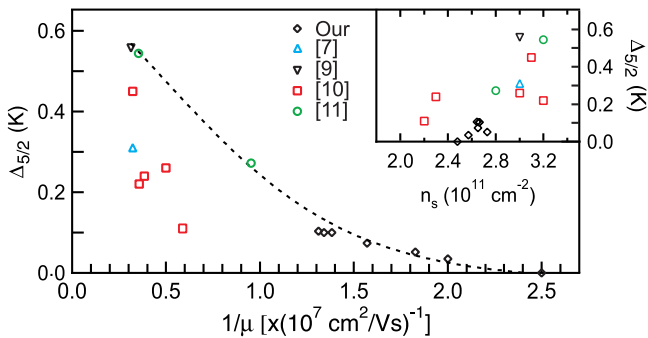}
\caption{(color online). Comparison between $\Delta_{5/2}$ of our samples
and values reported in the literature, plotted versus inverse
mobility $\mu^{-1}$ (main panel) and $n_{s}$ (inset). Dashed line
is a guide indicating the maximum $\Delta_{5/2}$ at each $\mu^{-1}$.}
\label{fig4}
\end{figure}

Effects of self-screening are also clearly visible for the $N=2$ LL; 
stripe and bubble phases~\cite{stripe-bubble} are well
developed for $N_{\mathrm{Si}}=0.5N_{0}$ but not for $0.15N_{0}$. Although 
not as dramatic, effects on the $N=0$ LL are also discernible. 

The screening of RI potential can be further enhanced by reducing $x$ from
$0.25$ to $0.24$ while keeping $N_{\mathrm{Si}}=0.5N_{0}$ [Fig.~3(c)]. This is
manifested by the better developed re-entrant insulating state and the stripe
phase showing a larger anisotropy. Despite these clear improvements in the
$R_{xx}$ features, activation measurements detected no significant change in
$\Delta_{5/2}$ ($=100$~mK). This presents the intriguing question as to whether
there should be a precise correspondence between the quality of the $R_{xx}$
features and the size of $\Delta_{5/2}$.

In Fig.\thinspace4, we compare the measured $\Delta_{5/2}$ of our samples 
with those reported in the literature for 30-nm QWs \cite{Eisenstein 2002, Kumar,
Pan2,Choi,Xia}. The inset plots the
data as a function of $n_{s}$, a measure of the Coulomb energy, showing that
the sizes of $\Delta_{5/2}$ in our samples are much smaller than previously 
reported for similar $n_{s}$. As shown in Refs. \cite{Dean,Pan2}, plotting $\Delta_{5/2}$
as a function of $\mu^{-1}$ makes clear that $\Delta_{5/2}$ tends to decrease with
decreasing $\mu$, i.e., with increasing BI concentration. Here, two important observations
are noteworthy. First, Fig. 4 does not include pseudogap data \cite{Gammel,Pan3}, 
demonstrating that a well-developed $5/2$ state with activation
behavior can be observed in the $\mu$ range much lower than anticipated from the
previous analysis \cite{Pan2}. Our data, in turn, indicate that the $\mu^{-1}$ dependence of 
$\Delta_{5/2}$ is more gentle than the linear decrease expected from the simple 
lifetime broadening argument. Second, in contrast to the dramatic change in the quality 
of the $R_{xx}$ features
with $x$, the variation of $\Delta_{5/2}$ in our samples is rather small and seems to be governed
by the values of $\mu$ instead. This can be understood by noting that in our samples a large part 
(more than $80\%$) of the intrinsic gap is taken away by disorder due to BIs, which masks
the influence of RIs on $\Delta_{5/2}$. These observations elucidate the different roles played
by the two dominant sources of disorder; while the emergence of a fully developed $\nu=5/2$ state
is dictated by the RI potential, at $\mu<10^{7}$ cm$^{2}$/Vs the size of the gap is mostly 
limited by the BI potential. In turn, the scattering of $\Delta_{5/2}$ reported in the ultra-high-$\mu$
regime, where the influence of BIs is considered to be small, may be due to the different levels of
RI screening, which can arise from details of the doping schemes or illumination procedure. This implies
that the total gap reduction is not given by a simple sum of the individual contributions from different
sources of disorder. These insights, combined with the improvements
in the mobility and sample design, will help
to further enhance the stability of the $\nu=5/2$ state.

We thank M.~Yamaguchi for the photoluminescence characterization of some
samples, N.~Kumada and L.~Tiemann for discussions and technical assistance.


\begin{thebibliography}{99}
\bibitem {Negative-U}The charge state of DX centers is an issue of long
standing debate~\cite{Mooney}, and we will not go into its detail.
Nevertheless, we believe that our results cast new light on the donor state in
Al$_{x}$Ga$_{1-x}$As ($0.22<x<0.4$).
\bibitem {Cluster}Near the onset of parallel conduction, donor wave functions
may overlap to form locally conducting clusters, which do not contribute to transport but
can provide screening. The fact that the $x=0.25$ samples with
$N_{\operatorname{Si}}=N_{0}$ and $0.5N_{0}$ show $\nu=5/2$ states of equal
quality indicates that the formation of such clusters is not essential.
\bibitem {Extra doping}In these samples, additional doping with
$N_{\operatorname{Si}}=0.25N_{0}$ was incorporated on each side at another
$60$-nm\textrm{ }setback to compensate unintentional trap states
and ensure the same $n_{s}$.
\bibitem {Correlation}The fact that the visibility of the $\nu=5/2$ state in
the $x=0.30$ sample with much higher $N_{\operatorname{Si}}=2N_{0}$ is similar
to that in the $N_{\operatorname{Si}}=0.15N_{0}$ sample (with $x=0.25$) seems to suggest
that such correlation is of secondary importance. Yet, considering the different
freeze out temperature for distinct $x$, we cannot entirely exclude the effect.
\end{thebibliography}
\end{document}